\documentstyle[emulateapj,apjfonts,psfig]{article}

\def\gs{\mathrel{\raise0.35ex\hbox{$\scriptstyle >$}\kern-0.6em 
\lower0.40ex\hbox{{$\scriptstyle \sim$}}}}
\def\ls{\mathrel{\raise0.35ex\hbox{$\scriptstyle <$}\kern-0.6em 
\lower0.40ex\hbox{{$\scriptstyle \sim$}}}}

\lefthead{Pearce et al.}
\righthead{Cosmological galaxy formation}

\def\P3M{\hbox{$P^{3}M$}}

\def\AP3M{\hbox{$AdP^{3}M$}}

\def\cc2{c2}
\def\cc3{c3}
\def\cc4{c4}
\def\cc{c}

\def\circ{circular}

\def\ApJ{ApJ}

\def\ApJS{ApJS}

\def\MN{MNRAS}

\def\etal{{\it et al.\thinspace}}

\def\spose#1{\hbox to 0pt{#1\hss}}
\def\approxlt{\mathrel{\spose{\lower 3pt\hbox{$\sim$}}
	\raise 2.0pt\hbox{$<$}}}
\def\approxgt{\mathrel{\spose{\lower 3pt\hbox{$\sim$}}
	\raise 2.0pt\hbox{$>$}}}

\def\<{\thinspace}

\def\boxit#1{\vbox{\hrule\hbox{\vrule\kern3pt\vbox{\kern3pt
          #1 \kern3pt}\kern3pt\vrule}\hrule}}

\def\kpc{{\rm\thinspace kpc}}

\def\Mpc{{\rm\thinspace Mpc}}
\def\hmpc{{$h^{-1}\rm\thinspace Mpc$\ }}
\def\Msun{\hbox{$\rm\thinspace M_{\odot}$}}

\def\h50{\hbox{$\rm\thinspace h_{50}$}}
\def\h50m1{\hbox{$\rm\thinspace h_{50}^{-1}$}}

\begin{document}
\title{A simulation of galaxy formation and clustering}

\author{F. R. Pearce\altaffilmark{1}, A. Jenkins\altaffilmark{1}, 
C. S. Frenk\altaffilmark{1}, J. M. Colberg\altaffilmark{2},
S. D. M. White\altaffilmark{2}, P. A. Thomas\altaffilmark{3}, 
H. M. P. Couchman\altaffilmark{4}, J. A. Peacock\altaffilmark{5}
and G. Efstathiou\altaffilmark{6} (The Virgo Consortium)}

\altaffiltext{1}{Physics Department, South Rd, Durham, DH1 3LE, UK}
\altaffiltext{2}{Max-Plank-Institut fur Astrophysik, Garching bei
Munchen, D-85740, Germany}
\altaffiltext{3}{Astronomy Centre, CPES, University of Sussex, Falmer, Brighton, BN1 9QJ, UK}
\altaffiltext{4}{Department of Physics and Astronomy, Univ. of Western
Ontario, London, Ontario, N6A 3K7, Canada}
\altaffiltext{5}{Institute for Astronomy, University of Edinburgh, 
Royal Observatory, Edinburgh EH9 3HJ, UK}
\altaffiltext{6}{Institute of Astronomy, Madingley Rd, Cambridge, CB2
0HA, UK}

\begin{abstract}

We discuss early results from the first large N-body/hydrodynamical
simulation to resolve the formation of galaxies in a cold dark matter
universe. The simulation follows the formation of galaxies by gas cooling
within dark halos of mass a few times $10^{11}\Msun$ and above, in a flat
universe with a positive cosmological constant. Over 2200 galaxies form in
our simulated volume of $(100 \Mpc)^3$. Assigning luminosities to the model
galaxies using a spectral population synthesis model results in a K-band
luminosity function in excellent agreement with observations. The two-point
correlation function of galaxies in the simulation evolves very little
since $z=3$ and has a shape close to a power-law over four orders of
magnitude in amplitude. At the present day, the galaxy correlation function
in the simulation is antibiased relative to the mass on small scales and
unbiased on large scales. It provides a reasonable match to observations.

\end{abstract}


\section{Introduction}

Studies of galaxy formation have advanced at an unprecedented rate in the
past few years. Data from the Keck and Hubble Space telescopes have
revolutionised our view of the high redshift Universe (e.g. Steidel \etal
1998) and have led to claims that the main phases of galaxy formation
activity may have now been observed (Baugh \etal 1998). From the
theoretical point of view, modelling galaxy formation presents a formidable
challenge because it involves the synthesis of a wide range of disciplines,
from early universe cosmology to the microphysics and chemistry of star
formation.


Because of the strongly non-linear and asymmetric nature of gravitational
collapse, the problem of galaxy formation is best addressed by direct
numerical simulation. The main difficulty stems from the huge range of
scales spanned by the relevant processes, from star formation to
large-scale clustering, which cannot all be simultaneously resolved with
current simulation techniques. Two complementary strategies have been
developed to deal with processes occurring below the resolution limit of a
simulation. In one of them, a semi-analytic model of the dynamics of gas
and star formation is used, for example, in conjunction with N-body
simulations of the formation of dark matter halos (Kauffmann \etal 1997,
1999a,b; Benson \etal 1999). This technique permits a large dynamic range
to be followed, at the expense of simplifying assumptions, such as
spherical symmetry, for the treatment of the dynamics of cooling gas and
star formation. The alternative approach, which is the one adopted in this
paper, is to solve directly the evolution equations for gravitationally
coupled dark matter and dissipative gas. This enables the dynamics of the
gas to be treated with a minimum of assumptions, at the expense of a severe
reduction in the accessible dynamic range. As in the semi-analytic
approach, a phenomenological model for star formation and feedback is
required.

Eulerian and Lagrangian numerical hydrodynamics have been used to simulate
galaxy formation. At present only the latter, implemented by means of the
Smooth Particle Hydrodynamics or SPH technique, provides sufficient
resolution to follow the formation of individual galaxies. For example, the
best Eulerian simulations to date, such as those of Blanton \etal (1999),
have gas resolution elements of $\sim 300-500$ kpc, whereas the early SPH
simulation of Carlberg \etal (1990) had a spatial resolution of 20 kpc. This, 
together with the simulations of Katz \etal (1992), Evrard \etal (1994) and 
Frenk \etal (1996), were the
first to resolve individual galaxies in relatively large volumes, allowing
detailed studies of the distribution of galaxies and the dynamics of
galaxies in clusters. However, the volumes modelled in this early work were
much too small to allow reliable investigations of galaxy clustering at the
present day. Galaxy clustering at high redshift has been investigated in
the simulations of Katz \etal (1999).

In this letter we present the first results of a large N-body/SPH
simulation of galaxy formation in a representative volume of a cold dark
matter universe, employing about an order of magnitude more particles than
the largest previous study of this kind (Katz \etal 1996). Our
simulation produced 2266 galaxies at the present day, compared to 60
in the simulation of Katz \etal (1992) and 58 in those of Katz
\etal (1996), the only other large SPH calculations to have been evolved
to the present.
Earlier dark matter simulations by the Virgo Consortium (Jenkins \etal
1998) demonstrated the kind of biases required for CDM universes to provide
a good match to observations of galaxy clustering. Here we show that these
biases arise quite naturally.

\section{The simulation}

We have simulated a region of a CDM universe with the same cosmological
parameters as the $\Lambda$CDM simulation of Jenkins \etal (1998): mean
mass density parameter, $\Omega_0=0.3$; cosmological constant,
$\Lambda/(3H_0^2)=0.7$; Hubble constant (in units of 100 km s$^{-1}$
Mpc$^{-1}$), $h=0.7$ (hereafter we adopt this value, unless otherwise
stated), and rms linear fluctuation amplitude in 8\hmpc spheres,
$\sigma_8=0.9$.  The baryon fraction was set to $\Omega_bh^2=0.015$, from
Big Bang nucleosynthesis constraints (Copi \etal 1995). We assumed an
unevolving gas metallicity of 0.3 times the solar value. The simulation was
carried out using ``parallel Hydra'', an adaptive, particle-particle,
particle-mesh N-body/SPH code (Pearce \& Couchman 1998), based on the
publicly released serial version of Couchman \etal (1995).

Our simulation followed 2097152 dark matter particles and the same number
of gas particles in a cube of side $100 \Mpc$ and required 12492 
timesteps to evolve from $z=50$ to $z=0$. The gas mass per particle is
$\sim 2\times10^9\Msun$ and, since we typically smooth over 32 SPH
neighbors, the smallest resolved objects have a gas mass of
$6.4\times10^{10}\Msun$. We employed a comoving $\beta$-spline
gravitational softening, equivalent to a Plummer softening of $14.3 \kpc$
until z=2.5. Thereafter, the softening remained fixed at this value, in
physical coordinates, and the minimum SPH resolution was set to match
this value. With our chosen parameters, our simulation was able to follow
the cooling of gas into galactic dark matter halos. The resulting
``galaxies'' typically have 50-1000 particles. With a spatial resolution of
$14.3 \kpc$, we cannot resolve the internal structure of galaxies and we
must be cautious about the possibility of enhanced tidal disruption, drag,
and merging within the largest clusters. However, as we argue below, there
is no evidence that this is a major problem.

As in all studies of this type, a phenomenological model is required to treat
physical processes occurring below the resolution limit of the simulation.
The first of these is the runaway cooling instability present in hierarchical
clustering models of galaxy formation. At high redshift, the cooling time
in dense subgalactic objects is so short that most of the gas would cool
(and presumably turn into stars) unless other processes acted to counteract
cooling (White \& Rees 1978, Cole 1991, White \& Frenk 1991).  
Since all the gas in the universe has clearly 
not cooled into dark matter halos, a common
assumption is that feedback from early generations of stars will
have reheated the gas, preventing it from cooling catastrophically.

Although a variety of prescriptions have been used to model feedback
(e.g. Navarro \& White 1993, Steinmetz \& M\"uller 1995, Katz \etal 1996), this process
remains poorly understood. In cosmological SPH simulations, gas can only
cool efficiently in objects above the minimum resolved gas mass, in our
case $6.4\times 10^{10}\Msun$. Thus, resolution effects alone act as a
crude form of feedback. Semi-analytical models of galaxy formation suggest
that feedback is relatively unimportant on mass scales above our resolution
limit.
We do not, therefore, impose any prescription for feedback over and above
that provided naturally by resolution effects. If the rate at which gas
cools in the simulation is identified with the rate at which stars form,
our adopted parameters give rise to a cosmic history of star formation in
broad agreement with data from $z\simeq 4$ to the present (Madau \etal
1998).

The second sub-resolution process that we must model is star formation and
the associated coupling of different gas phases in the interstellar
medium. Like feedback, this is a complex and poorly understood phenomenon.
In some SPH simulations, groups of cooled gas particles have been
identified with galaxies (the ``globs'' of Evrard \etal 1994). One
disadvantage of this procedure is that dense knots of cold gas can affect
the cooling of surrounding hot material because of the smoothing inherent
in the SPH technique. To avoid this problem, an alternative strategy often
used is to assume that gas that has cooled turns into collisionless
``stars'' according to some heuristic algorithm (Navarro \& White 1993, Katz
\etal 1996, Steinmetz \& M\"uller 1995).  This prescription effectively decouples the
cooled gas from the hot component. A drawback is that ``stellar'' systems
made up of only a few particles are fragile and can easily be disrupted.

We have adopted an intermediate strategy intended as a compromise between
the extremes of letting clumps of cool gas persist in the simulation and
turning them into stars. As in the first case, we identify galaxies with
groups of gas particles that have cooled below $10^4 {\rm K}$. However,
when computing the SPH density of particles with temperatures above $10^5
{\rm K}$, we do not include any contribution from particles below 
$10^4 {\rm K}$. All other SPH interactions remain unaffected. As in the case
where cool gas is turned into stars, our model effectively decouples the
galactic material from the surrounding hot halo gas, but unlike this case,
``galaxies'' in our model are made of dissipative material and thus are more
resilient to tidal interactions and mergers than model stellar
galaxies. Our model of the intergalactic medium can be regarded as a
simple, first step towards a multiphase implementation of SPH, an important
requirement when dealing with situations in which there are steep density
gradients.
The main effect of our treatment of cool gas is to prevent the formation of
very massive galaxies in the centers of the richest clusters in the
simulation, as happened, for example, in the simulation of Frenk \etal
(1996). Thacker \etal (1998) present a more detailed discussion of the
effects of runaway cooling and the production of supermassive objects.

\hbox{~}
\vskip 2mm
\centerline{\psfig{file=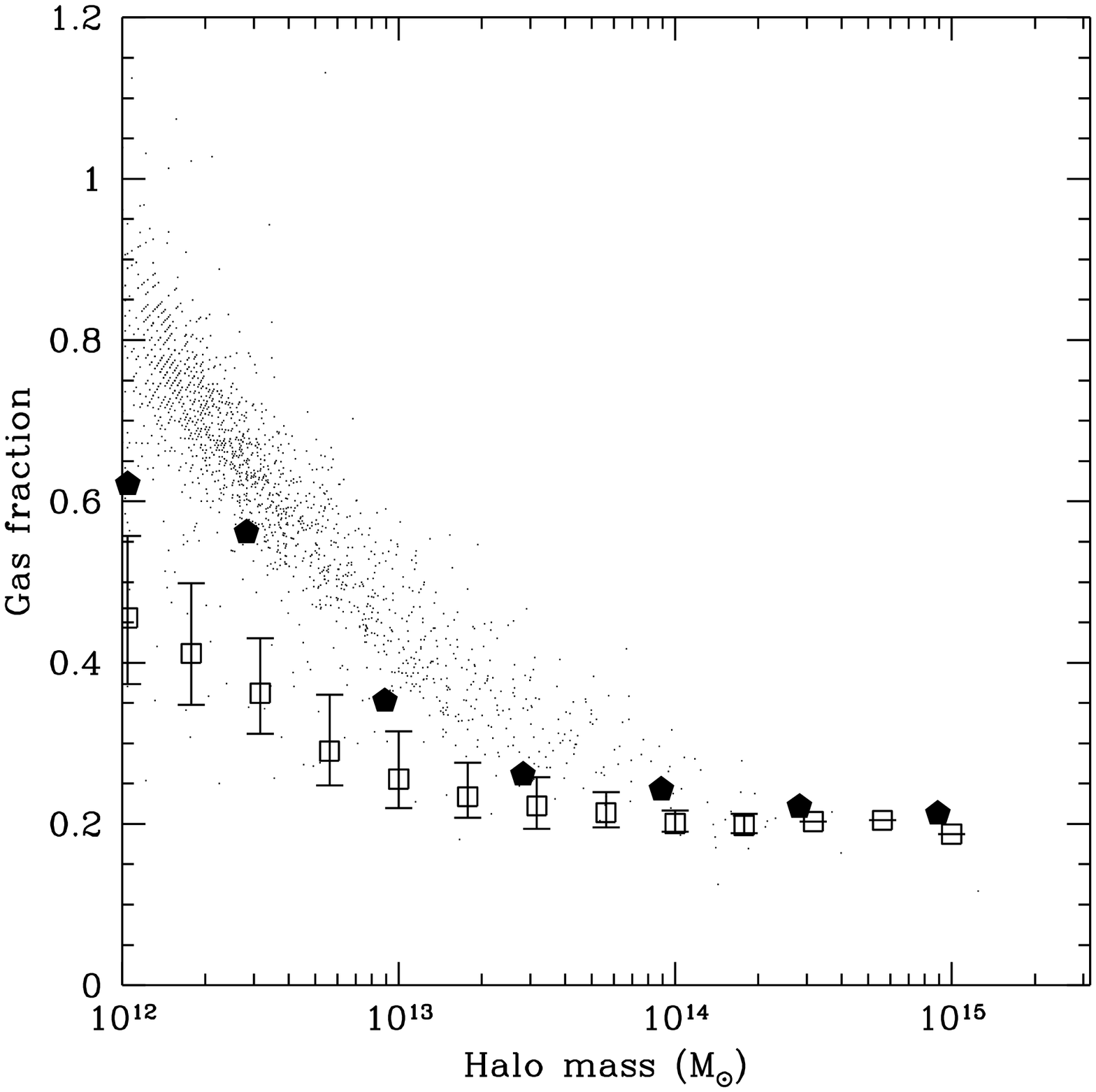,height=8.5cm}}
\noindent{
\scriptsize \addtolength{\baselineskip}{-3pt} 
{\bf Fig.~1.}  The mass fraction in the form of cold gas in virialized
halos, normalized to the average baryon fraction in the simulation. The
dots show results from the N-body/SPH simulation. The pentagons show
results from a semi-analytic model constrained to match the parameters of
the simulation.  The open squares show results from a full semi-analytic
model. 
\vskip 2mm
\addtolength{\baselineskip}{3pt}
}

As a test of our techniques, Fig. 1 compares the amount of cold gas in
halos in our simulation with that predicted by the semi-analytic model
of Cole \etal (1999). Halos in the simulation were located
by first identifying suitable centers using the friends-of-friends
group finder of Davis \etal (1985) with linking parameter, $b=0.05$,
and then growing spheres around these centers out to the virial radius
(defined as the radius within which the mean overdensity is 323; see
Eke \etal 1996). Only halos with more than 50 dark matter particles
were considered, of which there are 2353 in the simulation, spanning
over 3 orders of magnitude in mass. The simulation results (shown as
dots in the figure) are compared with two versions of the
semi-analytic model. In the first (filled pentagons), the parameters
of the semi-analytic model were set so as to mimic the conditions in
the simulation as closely as possible. The mass resolution of the
semi-analytic model was degraded to that of the simulation and the
feedback was switched off. The agreement between the simulation and
modified semi-analytic model indicates that there are only minor
differences in the cooling properties of the gas calculated with these
two different techniques. The second comparison is intended as a test
of how well the artificial ``feedback'' produced by resolution effects
compares with the physically motivated feedback prescription used in
semi-analytic models. In this case (open squares), the parameters of
the semi-analytic model were set so as to obtain a good match to the
faint end of the galaxy luminosity function, as discussed by Cole
\etal (1999). The agreement with the simulation in this case is
moderate.  Clearly, feedback in the semi-analytic model prevents
cooling within galaxy halos more efficiently than do resolution
effects in the simulation, but the difference is only about 50\% for
halos similar to that of the Milky Way.

\hbox{~}
\vskip 2mm
\centerline{\psfig{file=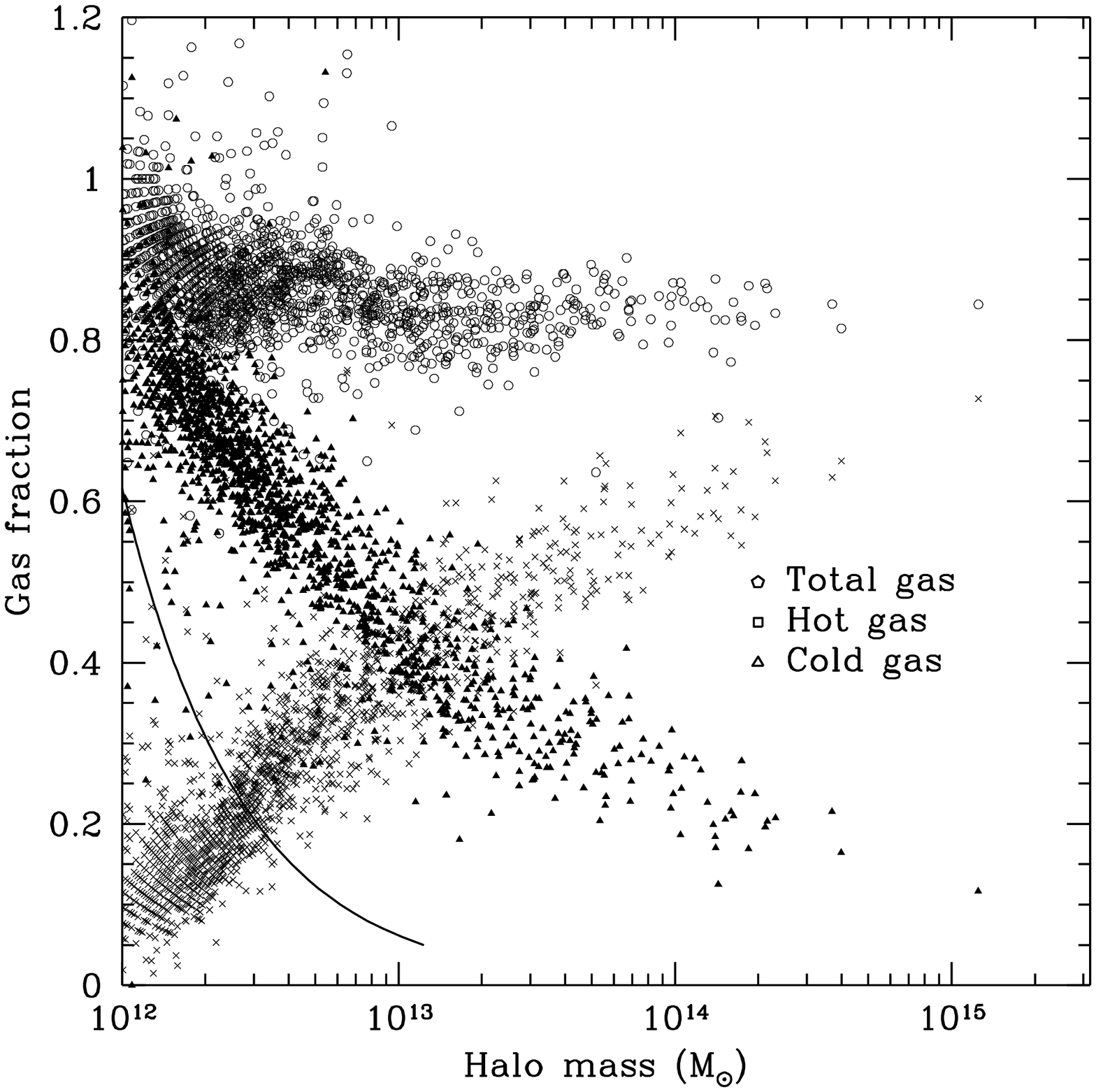,height=8.5cm}}
\noindent{
\scriptsize \addtolength{\baselineskip}{-3pt} 
{\bf \noindent Fig.~2.} The ratio of gas mass to dark matter within the
virial radius of each halo, in units of the mean baryon fraction in the
simulation. The circles show the total gas mass fraction while the
crosses and triangles show the mass fractions of gas hotter and cooler than
$12000 {\rm K}$ respectively. The solid line shows the resolution limit of 32
gas particles.
\vskip 2mm
\addtolength{\baselineskip}{3pt}
}

\section{Results}

The ability of gas to cool is a strong function of the mass of the host
halo. Fig.~2 shows the fraction of hot and cold gas within the virial radius
of each halo, normalized to the mean baryon fraction of the simulation
(10\%), as a function of halo mass.  In small halos just above the resolution
limit (indicated by the solid line), most of the gas cools, but the
fraction of cold gas decreases rapidly with halo mass as the cooling time
for the hot gas increases. In large halos, most of the gas never cools. The
crossover point occurs at a halo mass of $\sim 10^{13} \Msun$. 
Because of the generally asymmetric and chaotic nature of halo
formation, a few low mass halos have baryon fractions in excess of the
universal mean, but in most galactic halos the baryon fraction ranges
between 80 and 100\% of the cosmic mean. On the scale of galaxy clusters,
the baryon fraction is 85\%, similar to the values obtained by White \etal
(1993) and Frenk \etal (1999).


We identify ``galaxies'' in our simulation with dense knots of cold
gas. These are very easy to locate except in the minority of cases where a
merger is ongoing or where the galaxy is experiencing significant tidal
disruption or ablation within a cluster halo. To find galaxies we used the
friends-of-friends group finder, with a linking length of $0.0164(1+z)$
times the mean comoving interparticle separation.  This selects material
with overdensity greater than $\sim 10^5$ at $z=0$.  The galaxy catalogue
is almost unaffected by large changes in the maximum linking length. At the
end of the simulation there were 2266 resolved objects within the volume.

\hbox{~}
\vskip 2mm
\centerline{\psfig{file=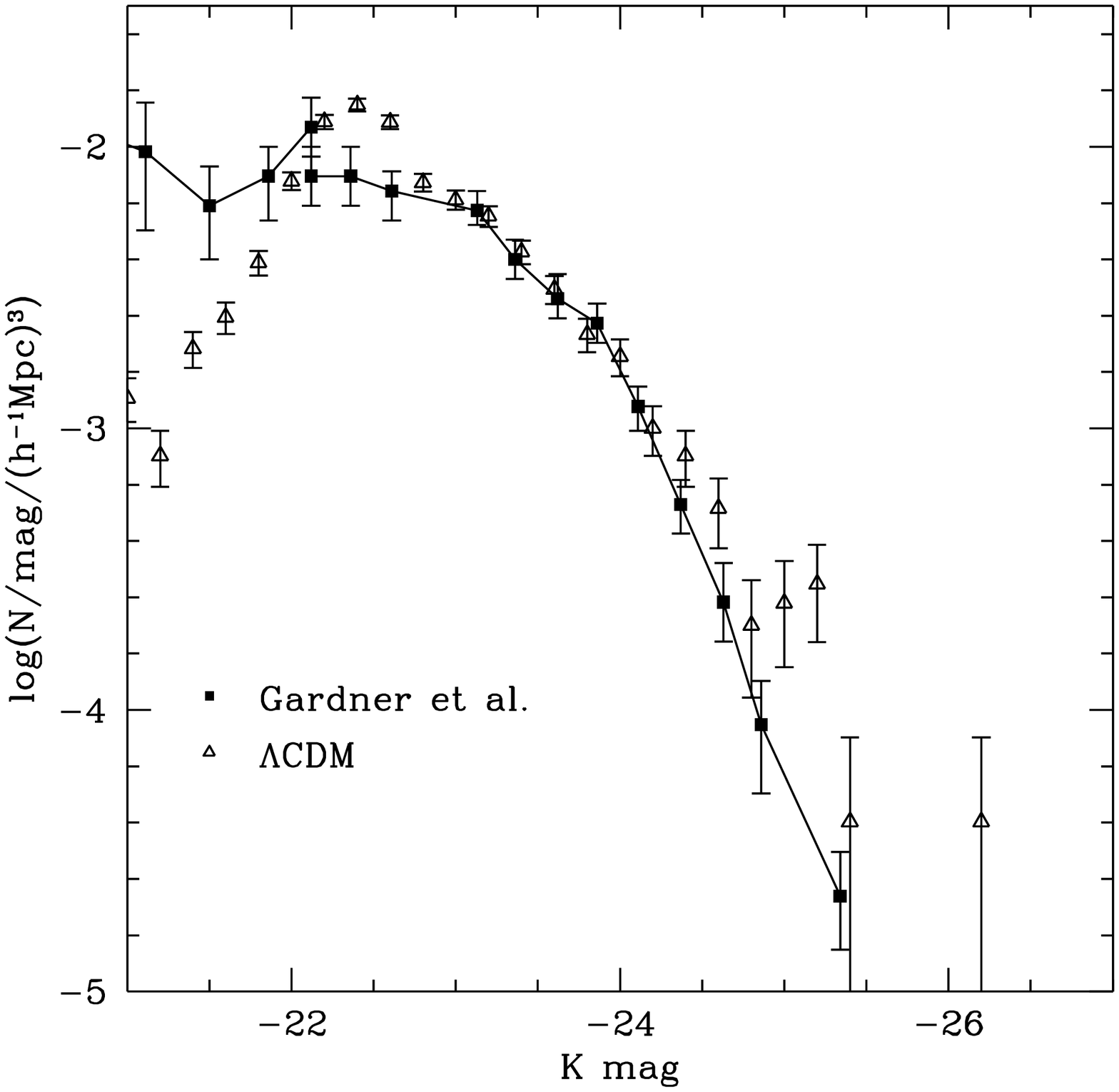,height=8.5cm}}
\noindent{
\scriptsize \addtolength{\baselineskip}{-3pt} 
{\bf Fig.~3.}  A comparison between the K-band galaxy luminosity function
in the simulation with observations. The simulation data are shown by open
triangles and the data from Gardner \etal (1997) by filled squares. A
luminosity normalization factor of $\Upsilon=2.8$ has been assumed. Poisson
errors are shown.
\label{kcount}
\vskip 2mm
\addtolength{\baselineskip}{3pt}
}

We can assign a luminosity to each galaxy in our simulation using the
stellar population synthesis model of Bruzual \& Charlot (1993). For this
purpose we assume that at each model output, a fraction, $1/\Upsilon$, of
the gas that has cooled since the previous output turns into stars in an
instantaneous burst with a Salpeter IMF whose subsequent spectrophotometric
evolution is given by the synthesis model. Because the output times were
relatively infrequent, this procedure works best for K-band luminosities.

In Fig.~3 we compare the resulting K-band galaxy luminosity function with
the observational data of Gardner \etal (1997). The shape of the model
luminosity function agrees well with the data, and the model and observed
functions match well if we set the luminosity normalization factor,
$\Upsilon =2.8$. This implies that only 35\% of the cold gas has been
turned into visible stars, with the rest remaining in dense gas clouds and
brown dwarfs or hidden in some other form. The associated mass-to-light
ratios are about twice as large as those measured for elliptical galaxies,
but these numbers should be treated with caution. Not only is our star
formation prescription very crude, but our model ignores the effects of
metallicity and obscuration by dust. Furthermore, as the comparison with
the full semi-analytic model indicates, too much gas has probably cooled in
our simulation because of our neglect of feedback processes. In spite of
these reservations, the agreement in Fig.~3 is very good and suggests that
our simulation provides a realistic description of the formation of bright
galaxies.

The relatively large volume of our simulation allows a reliable measurement
of the clustering properties of galaxies and their relation to the
clustering properties of the mass. The galaxy and mass two-point
correlation functions at various epochs are plotted in Fig.~4.  The mass
correlation function agrees very well with the results of our earlier, dark
matter only simulations which followed a cubic region of side $342\Mpc$
using 16.8~million particles (Jenkins \etal 1998). The clustering amplitude
of the mass grows by a factor of about 30 between the two epochs shown in
the figure, $z=3$ and $z=0$. By contrast, the galaxy correlation function
hardly evolves at all between $z=3$ and $z=0$.

\hbox{~}
\vskip 2mm
\centerline{\psfig{file=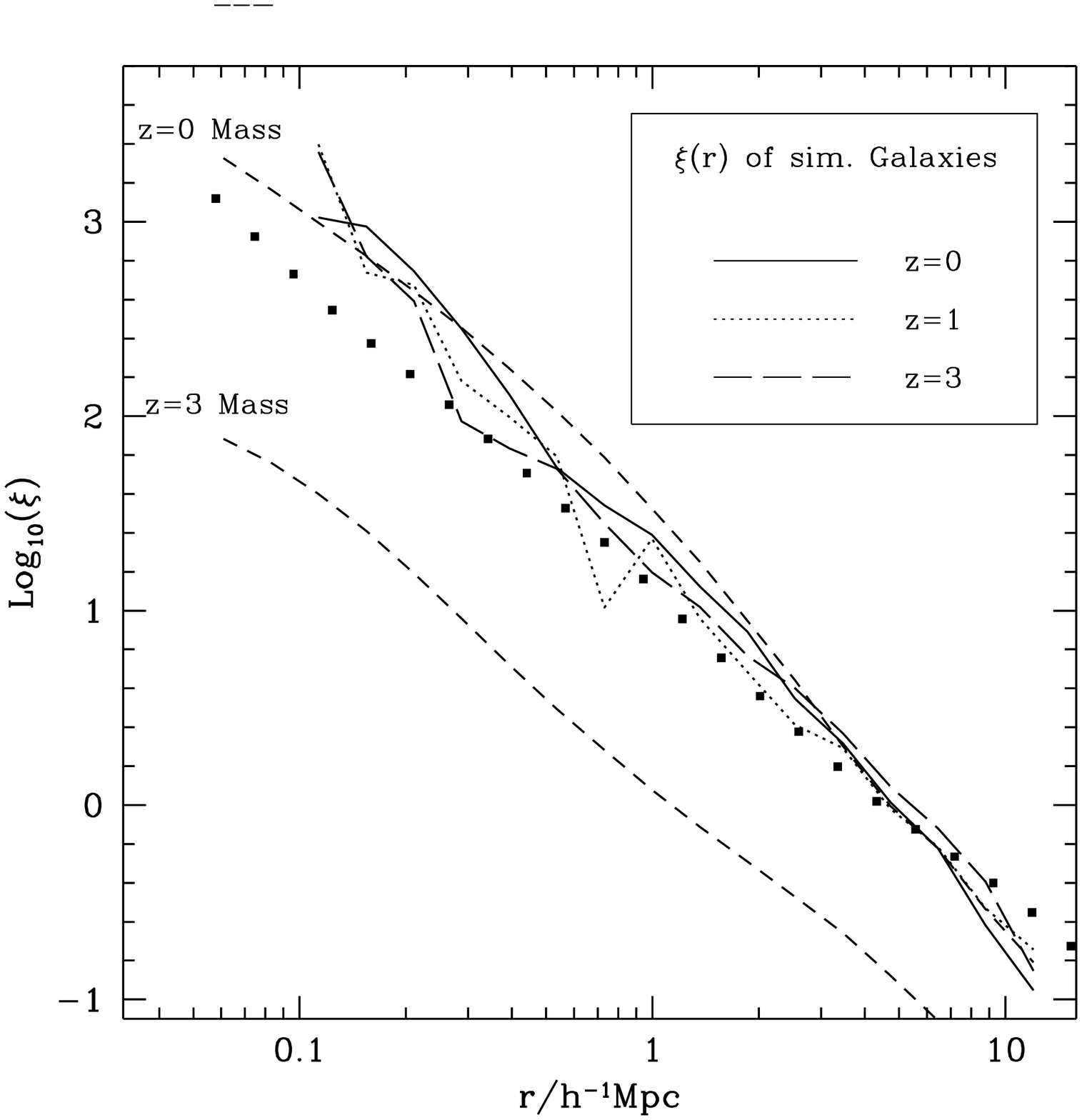,height=8.5cm}}
\noindent{
\scriptsize \addtolength{\baselineskip}{-3pt} 
{\bf Fig.~4.}  Mass and galaxy correlation functions. The dashed lines show
the mass correlation functions at $z=3$ and $z=0$. The solid, long dashed
and dotted lines show the galaxy correlation functions in the simulation at
the indicated redshifts. The squares show the observed, real-space
correlation function, estimated by Baugh (1996), from the APM survey.
\vskip 2mm
\addtolength{\baselineskip}{3pt}
}

The difference between the clustering growth rates of galaxies and mass is
a manifestation of ``biased galaxy formation'', the preferential formation
of galaxies in high peaks of the primordial density field. It was already
seen in the first simulations of cold dark matter models by Davis \etal
(1985), in which galaxies were put in "by hand" near high peaks of the
initial density field.  It is also very clear in the SPH simulations of
Evrard \etal (1994) and Katz \etal (1999), and can even be inferred from
the N-body only simulations of Bagla (1998) and Colin \etal (1999).
Semi-analytic techniques have been used on their own (Baugh \etal 1999), or
combined with N-body simulations (Kauffmann \etal 1999b), for detailed
study of the clustering evolution of galaxies, while specific applications
to high redshift Lyman-break galaxies are to be found in Baugh \etal (1998)
and Governato \etal (1998). The latter models provide an excellent
description of the strong clustering discovered by Adelberger \etal (1998).

In Fig.~4 we also plot the observed, real-space galaxy correlation function
at $z\simeq 0$, estimated by Baugh (1996) from the APM survey (filled
squares.) This may be compared with the $z=0$ results in our simulation
(solid line). On scales larger than a few hundred kpc the agreement is
good.  (The differences at $r\gs 10$\hmpc are due, for the most part, to
finite volume effects, as we have verified by comparison with the larger
simulations of Jenkins \etal 1998.) Beyond 1\hmpc the galaxy correlation
function is very close to the mass correlation function.  On smaller scales
galaxies are less strongly clustered than the mass, or antibiased, an
effect that persists until separations of $\sim 100 h^{-1}$kpc. At small
separations the model correlations lie above the APM data.  Over nearly
four orders of magnitude in amplitude, the model galaxy correlation
function is very close to a power-law even though the mass correlation
function is not. An essentially featureless galaxy correlation function was
also obtained for the same cosmological model in the semi-analytic model of
Benson \etal (1999) and, for some parameter combinations, in those of
Kauffmann \etal (1999a).

\section{Conclusions} 

The simulation presented here is the first to resolve galaxy formation
in a large enough volume to allow a reliable study of the demographics
and clustering of galaxies.  Our results are encouraging: the
resulting luminosity function and correlation function of galaxies are
broadly consistent with observations. Furthermore, the correlation
function of bright galaxies in the simulation changes little since
$z=3$, in agreement with results from semi-analytic studies and with
available data at high redshift. Further progress will require a more
detailed treatment of the astrophysics of galaxy formation,
particularly of the processes of star formation and feedback.

\section*{Acknowledgments}

This work was carried out as part of the programme of the Virgo Consortium,
using the facilities of the Computing Centre of the Max-Planck Society in
Garching and the Edinburgh Parallel Computing Centre.  FRP, PAT and HMPC
acknowledge a NATO collaborative research grant (CRG 970081). This work was
supported by the EC network for ``Galaxy formation and evolution.''  We
thank Carlton Baugh and Shaun Cole for providing unpublished results from
their semi-analytic models.

\end{document}